\documentclass{ledger}
\usepackage{xcolor}

\usepackage{cleveref}
\usepackage{array}
\usepackage{booktabs}
\usepackage{tabularx}

\newcommand{\thefirstpagenum}[0]{1}
\newcommand{\thelastpagenum}[0]{5}
\newcommand{\theyear}[0]{201X}
\newcommand{\thevol}[0]{X}
\newcommand{\thedoi}[0]{DOI 10.5195/LEDGER.\theyear.X}
\newcommand{\ledgerpages}[0]{\thefirstpagenum-\thelastpagenum}

\hypersetup{pdfauthor={First Last}, pdftitle={This is a Title}}

\newcommand{\appref}[1]{Appendix~\ref{#1}}

\title{Towards an Optimal Staking Design: Balancing Security, User Growth, and Token Appreciation}
\author{Nicolas~Oderbolz\and Beatrix Marosvölgyi\and\thanks{N.~Oderbolz (nicolas.g@cryptecon.org) and B. Marosvölgyi (beatrix@cryptecon.org) are Economists at the Center for Cryptoeconomics, Zurich, Switzerland.}  Matthias~Hafner\thanks{M. Hafner (matthias@cryptecon.org) is the Director of the Center for Cryptoeconomics, Zurich, Switzerland.}}

\pretitle{
  \centering 
  \fontsize{24pt}{28pt}\selectfont} 

\usepackage{hyperref}
\usepackage{makecell}

\begin{document}

\maketitle

\thispagestyle{pagefirst}

\begin{abstract} This paper examines the economic and security implications of Proof-of-Stake (POS) designs, providing a survey of POS design choices and their underlying economic principles in prominent POS-blockchains. The paper argues that POS-blockchains are essentially platforms that connect three groups of agents: users, validators, and investors. To meet the needs of these groups, blockchains must balance trade-offs between security, user adoption, and investment into the protocol. We focus on the security aspect and identify two different strategies: increasing the quality of validators (static security) vs. increasing the quantity of stakes (dynamic security). We argue that quality comes at the cost of quantity, identifying a trade-off between the two strategies when designing POS systems. We test our qualitative findings using panel analysis on collected data. The analysis indicates that enhancing the quality of the validator set through security measures like slashing and minimum staking amounts may decrease dynamic security. Further, the analysis reveals a strategic divergence among blockchains, highlighting the absence of a single, universally optimal staking design solution. The optimal design hinges upon a platform’s specific objectives and its developmental stage. This research compels blockchain developers to meticulously assess the trade-offs outlined in this paper when developing their staking designs.
\\

\end{abstract}

\section{Introduction} \label{sec:Intro}
Staking is an essential mechanism that underpins the security infrastructure of many blockchain networks. The staking process involves participants locking cryptoassets on a blockchain as collateral for honest behavior in exchange for a reward. Staking has evolved from a specialised activity for technology enthusiasts to a mainstream service accessible through traditional financial institutions. As a result, staking has seen massive growth. For instance, Ethereum, the most widely used Proof-of-Stake (POS) blockchain, has attracted more than USD 50 billion worth of staked ETH in just three years.\cite{Beaconcha} Due to the massive volume of assets locked into these protocols, as well as the central importance of the consensus mechanism to the security of the blockchains, it is critical to understand how stakeholders respond to different changes in the staking mechanisms. Staking is not unique to Ethereum, as many modern blockchains use variations of the mechanism. Typically, staking implementations differ in operational nuances. The result is a set of complex ecosystems where the identification of a single best design approach through comparative analysis remains elusive. The various divergences in the design of POS mechanisms can make it difficult to understand the overall impact of staking design choices on the network's staking behaviour and, as a result, its economic security.

Despite the importance of the topic, the academic literature investigating the impact of staking design choices on the staking behaviour is still small. The latest research on Proof-of-Stake blockchains has focused on different topics, such as detailing the different types of POS consensus mechanisms \cite{Saleh,Schaaf}, understanding specific security threats such as 51\% attacks \cite{Huang2021,Pierro&Tonelli} or double-spending \cite{Karpinski,Iqbal&Matulev}. Furthermore, many authors have explored the threats that asset centralisation poses to the integrity of the blockchain and have tried to quantify its magnitude \cite{Irresberger,Sai2021,Kogan2023}. Others have focused more on specific aspects of staking designs. For instance, \textit{Chitra (2021)} \cite{Chitra} argues that on-chain lending smart contracts may compromise platform security by discouraging staking in favour of token lending. \textit{Kose et al. (2021)} \cite{Kose2021} studied the effect of block rewards on the equilibrium staking level and showed that over a specific level, increasing block rewards can have a negative effect on the level of staking equilibrium. Further, \textit{Gersbach et al. (2022)} \cite{Gersbach} studied the formation of delegation pools in the presence of malicious agents. They found that there is a trade-off between improved returns for honest stakeholders and potential security risks. Their reasoning is that an optimal reward distribution mitigates malicious gains, but unregulated distribution leaves blockchains vulnerable, since malicious agents could gain more delegators by rewarding them more generously.

The latest working paper by \textit{Gogol et al. (2024)} \cite{Gogol} highlights another strain of literature, which analyses the performance of major liquid staking tokens against traditional staking on POS-blockchains. Finally, the strain of literature the present paper contributes to most closely analyses the different factors that influence the staking ratio on POS-blockchains. \textit{Cong et al. (2022)} \cite{Cong2022} demonstrated the significant interplay between staking, token pricing, and reward rates in cryptocurrency markets. \textit{Noh et al. (2023)} \cite{Noh2023} analysed the openness of 11 POS-blockchains based on five metrics related to decentralisation. We build on this literature and explore the economic and security implications of different staking designs. Compared to previous analyses, this paper provides a comprehensive assessment of factors affecting blockchain security and quantifies their impact on staking. 

The present research focuses on seven prominent POS-blockchains: Algorand, Avalanche, Cosmos, Cardano, Ethereum, Polkadot, and Solana. The paper first evaluates the differences in staking designs and reward distribution mechanisms across these POS-blockchains, shedding light on the key design dimensions of these POS systems as well as their peculiarities. The findings are synthesized into a broad discussion of the prevailing staking design choices and their implications for various stakeholders within the blockchain ecosystem. This includes an exploration of the critical challenge of balancing security, user growth, and token appreciation required to successfully sustain a POS-blockchain. In addition, the paper conducts a focused analysis of the economic security aspect of staking, revealing a strategic divergence among blockchain networks between static and dynamic security measures. We show that there is a trade-off between optimising the quality of the validator set (\textit{static security}) and expanding the set of validators (\textit{dynamic security}), and discuss how different blockchains deal with this trade-off. We find that Solana, Algorand, and Cardano place a strong emphasis on dynamic security, having low or no slashing, and minimal requirements for staking amounts and duration. On the contrary, Ethereum, Polkadot, and Cosmos impose stringent requirements in most areas, thus adopting a static security approach. Avalanche stands as an exception to this rule. 

Finally, we empirically demonstrate the existence of the trade-off between static and dynamic security using panel data analysis: factors that increase the quality of validators tend to decrease the amount staked relative to total supply. Our estimates indicate that, in particular, severe slashing conditions and higher minimum staking durations reduce the staking ratio.

The remainder of this paper is structured as follows. Section \ref{sec:Economics} introduces the basic economics of staking and the main concepts on which we build. Section \ref{sec:QualAssess} presents a qualitative assessment of staking factors, such as minimum staking amount, minimum staking periods, staking rewards, and slashing. Section \ref{sec:QuantAssess} presents a short quantitative assessment of the staking factors discussed, and Section \ref{sec:Conclu} concludes.

\section{Economics of Staking} \label{sec:Economics}
This section provides an overview of key elements of staking, the existing variations, and the objectives of POS-blockchains. This section also characterises POS-blockchains as economic platforms and examines the primary trade-offs.
\subsection{Main elements of staking}
Staking forms the foundation of the Proof-of-Stake consensus mechanism. Individuals lock their assets (typically native coins) on a blockchain, thereby securing the protocol.\cite{footnote0} The stake acts as a form of \textit{collateral} to ensure that so-called validators, who are responsible for verifying and appending the blockchain, act in an honest manner that is in line with the protocol's intentions. In the most basic variant of POS, only validators can stake (see also \appref{appx_a} for POS-system variations). As such, they are comparable to miners in a Proof-of-Work (POW) system, such as Bitcoin. Both validators and miners compete to contribute to the blockchain and in doing so to be \textit{rewarded} for their efforts. However, there is a difference in how they compete. While miners use computational resources to solve cryptographic puzzles in Proof-of-Work (POW), validators are chosen randomly by the consensus algorithm to propose a new block. In many cases, the probability of being selected depends on the size of the validator's stake. In this sense, there are similar incentives in the POS and POW consensus protocols: competition between validators is based on the size of their stakes, while competition between miners is based on computational power. The focus of investment in the protocol shifts from computational resources to native coins. Of course, it should be mentioned that although the computational requirements for solving cryptographic puzzles are not present in POS, validators still need to utilise some level of \textit{computational power} when they are selected for the proposal of the block. To incentivise this and to ensure that validators act truthfully, many POS-blockchains implement so-called \textit{slashing} penalties. This means that a failed or incorrect block proposal can result in the loss of the staking reward or even the collateral itself. In this manner, the reward for staking, which typically includes newly generated coins, can be viewed as compensation for the initial capital investment, the potential penalty for misconduct, and the expenses associated with supplying the computational power necessary for accurate transaction validation. \appref{appx_c} provides an overview of staking design mechanisms.

\subsection{Goals of POS-blockchains and platform economics}

When designing POS mechanisms, designers generally follow three goals: 
\begin{enumerate}
\item High security $\rightarrow$ high staking volume. Staking is essential for ensuring the security of a POS-blockchain. By locking up their assets, individuals are incentivised to validate transactions and maintain the integrity of the network. The more assets are locked in the blockchain as a stake, the more secure it becomes. Thus, the designers of the POS mechanism aim to achieve a higher staking volume.
\item Increased growth $\rightarrow$ ecosystem development and low fees. To achieve increased growth, blockchain designers must create a system that is attractive, efficient, and cost-effective. To grow, the platform can invest in development, marketing, and - if possible - ask for low transaction fees. Thus, to achieve platform growth, the designers of POS mechanisms aim to increase ecosystem development expenditures and keep transaction fees low.
\item High levels of investments $\rightarrow$ low levels of inflation. Low inflation is crucial for coin holders who want to see a return on their investment and avoid the dilution of their coins. Furthermore, inflation dilutes the rewards that stakers receive. Thus, POS mechanism designers ceteris paribus aim for low inflation.
\end{enumerate}

To achieve these goals, POS-blockchains need to connect different groups. In its simplest form, POS-blockchains can therefore be characterised as two-sided platforms that facilitate a connection between individuals who secure the blockchain (validators) and those who seek a network to transfer value without an intermediary (users). This dynamic can also be extended to a three-sided platform when considering investors who fund blockchain projects. The challenge of such multi-sided platforms is to balance the demands of the different user groups effectively. This balance is crucial for success.\cite{Rochet2003, Belleflame2021, Hafner2023}. Since POS blockchains cannot attract all three agent groups with the same effort, they face trade-offs. To enhance security, validators are needed that expect appropriate compensation for their efforts. In order to increased growth, users, developers, and liquidity providers need to be attracted. Finally, investors expect attractive token price performance, to compensate their substantial investments. Due to the associated costs, it is not feasible for the platform to simultaneously achieve all of these objectives. In analogy to the well-known blockchain trilemma, we describe these trade-offs as the Staking Trilemma.

A blockchain that aims to attract validators to the network in order to improve security will need to provide appropriate rewards to compensate the validators for their work and opportunity costs. The blockchain has two options to finance these rewards: increasing its coin supply or charging higher transaction fees to its users. To increase validator participation, the blockchain may need to accept either a decrease in user growth due to higher transaction costs or negative pressure on its coin price due to the increase in coin supply. Likewise, a blockchain that aims to attract users with low fees may need to either forgo attracting additional validators with high staking rewards or do so by employing an inflationary monetary policy, which in turn may be disadvantageous for investors. Lastly, a blockchain that prioritises the value of the coin and thus provides attractive returns for investors may likely need to forgo an inflationary monetary policy and thus either provide high staking rewards through increased fees or not incentivize validators in favor of retaining low fees. In sum, designers of POS-mechanisms face trade-offs and must carefully balance their priorities and implement strategies that support their long-term vision.

In addition to the above decisions (validator reward rate, inflation rate, and transaction fees), several other important policy parameters can be identified, especially with regard to the security of the platform. These factors include, but are not limited to, the methods and severity of punishment for bad validator behaviour (slashing), minimum staking amounts, and minimum staking durations. 

\cref{tbl:1} summarises the policy parameters set by the POS-blockchains mentioned earlier. Staking requirements and amounts, as well as approaches to slashing, vary significantly among different platforms. The minimum staking period requirements can range from no explicit requirements to a minimum duration of one month, and minimum staking amounts can range from zero to as high as USD 112'368. Similarly, approaches to slashing can vary from staking rewards being reduced to forfeiture of the entire stake.
\\
\begin{table} [h]
\caption{Overview of the staking mechanisms.}
\label{tbl:1}
\begin{tabularx}{\textwidth}{>{\raggedright\arraybackslash}X 
                              >{\centering\arraybackslash}X 
                              >{\centering\arraybackslash}X 
                              >{\centering\arraybackslash}X 
                              >{\centering\arraybackslash}X 
                              >{\centering\arraybackslash}X}
\hline
Blockchain & Average Reward Rate (pct.) & Average Inflation Rate (pct.) & Minimum Staking Period (days) & {Minimum Staking Amount}\footnote{Average USD value of minimum staking amount between 2021-2023.} & Slashing \\
\hline
Algorand & 8.5 & 2.5 & 0 & 0.1 ALGO \quad (0.1 USD) & Reduced rewards \\ 
Avalanche & 8.4 & 5.8 & 14 & 25 AVAX \quad (823 USD) & Reduced rewards \\ 
Cardano & 3.8 & 3.3 & 0 & 0 & Reduced rewards \\
Cosmos & 18.9 & 14 & 14 & 0 & Collateral slashing \\
Ethereum & 4.5 & 0.1 & 0 & 32 ETH (69'931 USD) & Collateral slashing \\
Solana & 6.3 & 7.3 & 2-3\footnote{Solana minimum duration is 1 epoch, which is roughly 2-3 days.} & 0 & Collateral slashing \\
Polkadot & 14.3 & 7.4 & 28 & 1 DOT \qquad (15 USD) & Collateral slashing \\
\hline
\end{tabularx}
\end{table}

Upon closer inspection, it is clear that with respect to security, two distinct strategies emerge among POS-blockchains. Some of them, such as Solana, Algorand, and Cardano, have low requirements across all dimensions, while others, including Ethereum, Polkadot, and Cosmos, impose high requirements in most of the discussed areas. Avalanche is an exception to this rule. In the following section, we will explore the topic of security in more detail to gain a better understanding of these distinctions.


\section{Qualitative Assessment of Staking Factors} \label{sec:QualAssess}
In this section, we discuss the economic trade-offs of different staking policy choices with respect to security. As mentioned above, the various dimensions of a staking policy can be combined in a variety of ways. For an overview of all the possible dimensions and their explanations, refer to \appref{appx_b}.
The goal of the following qualitative evaluation is not to identify a single best approach, but rather to provide an intuition for how individual parameters of staking programmes may affect the incentives faced by participants in a POS protocol. In particular, we will discuss the trade-offs involved in deciding on minimum stakes, minimum staking periods, staking rewards, and slashing.
\subsection{Minimum staking amount}
 In general, the \textit{minimum staking amount} influences a POS-blockchain along two dimensions: 1) static security and 2) dynamic security: 
\begin{enumerate}
\item Static security: Higher minimum staking requirements prevent validators with low levels of financial interest in the functioning of the protocol from participating in the consensus mechanism. Those that do participate will by design face stronger financial incentives to contribute to the proper functioning of the consensus mechanism, potentially improving its security.
\item Dynamic security: The lower the minimum staking requirements for validators, the lower the financial barriers for participation in the consensus mechanism. This can benefit validator adoption and may result in a larger and a more decentralized validator base.
\end{enumerate}

The minimum staking amount increases the static security but decreases dynamic security. The overall effect is therefore not straightforward, and the protocols must strike a balance between the two. Depending on the priorities of the protocol, different choices of minimum staking requirements may be reasonable. For example, newly established protocols may want to institute lower minimum staking requirements to foster validator adoption. On the other hand, established protocols may want to improve the quality of individual validators after they have reached a critical network size. In this case, increasing minimum staking requirements may be preferable. 
\subsection{Minimum staking period} 
act in a similar manner. They determine the minimum duration for which validators must stake their tokens when participating in the consensus mechanism. We can again identify a trade-off between (1) static security and (2) dynamic security:
\begin{enumerate}
\item Static security: Longer lock-up periods tend to increase the level of commitment to the protocol. With assets being committed for extended periods, potential reputation damages become more costly for individual validators. Furthermore, longer staking periods limit the ability to withdraw staked assets in response to short-term fluctuations in token value, aligning the financial interests of validators with the medium-term success of the token and platform. Additionally, extended staking periods can help alleviate the risk of substantial withdrawals from the staking pool occurring within a short period of time, akin to a “bank run”. Such events can potentially occur during valuation bubbles and can destabilize the consensus mechanism. \cite{Kogan2023} Therefore, given a static set of validators, extending the minimum staking period can help improve platform security. 

\item Dynamic security: As already mentioned, longer staking periods reduce the degree of flexibility with which stakers can reallocate their funds. This increases validators’ exposure to token price risk. Additionally, tokens are typically used as currency within the blockchain platform and as a result, higher minimum staking periods can reduce transaction and consumption convenience on the network. Minimum staking periods thus make the staking investment more prohibitive and increase the entry barriers to the staking program. Thus, by limiting validator adoption, longer staking periods decrease the security of the platform.  
\end{enumerate}
As with the minimum staking amount, the minimum staking period increases static security but reduces dynamic security, and protocols must strike a balance between the two. In practice, the trade-offs underlying the choice of minimum staking amount and minimum staking period are handled in a variety of ways. For example, Cosmos and Polkadot offer low minimum staking amounts but require at least some commitment in terms of staking duration. In comparison, Avalanche and Ethereum impose shorter lock-up periods but simultaneously expect substantial minimum staking amounts. Furthermore, a subset of blockchains, including Cardano and Solana, require both modest minimum staking periods and low minimum staking amounts. 

At present, blockchains handle the trade-off between minimum staking period and minimum staking amount differently. It may also be worth noting that until the Shapella update, validators on Ethereum could not unstake at all. This example shows that demand for staking can be sustained even with long minimum staking periods. The increased demand for staking after the Shapella update indicates, however, that the staking duration does play a crucial role in staking adoption. It is also worth noting that some blockchains, including Solana and Ethereum, mitigate the risk of bank runs by limiting the number of withdrawals that can be made in a given period of time.\cite{footnote} Such policies are able to ensure the stability of the active validator set during periods of high demand for withdrawals while allowing for flexibility in withdrawals when demand is stable.

\subsection{Staking Rewards}
We turn next to the rewards validators and delegators earn for staking and participating in the consensus mechanism. The staking reward policies analyzed in this paper exhibit significant variation across different blockchains. In particular, the choice of a staking rewards policy necessitates a careful consideration of the trade-offs among three key factors: (1) static security, (2) dynamic security, and (3) token success:

\begin{enumerate}
    \item Static security: High staking rewards and particularly staking rewards that depend on active participation in the consensus mechanism may improve the quality of the validator set. In addition, high staking rewards increase the financial costs for an adversary seeking to execute a bribery attack, since validators require higher compensation to forego their regular staking rewards.
    
    \item Dynamic security: All else equal, higher staking rewards serve as an incentive for participation in the consensus mechanism, resulting in a larger validator set and staking pool, thereby improving the security of the consensus mechanism. It is worth noting that improved protocol security can also have second-order effects on platform adoption and token price, as a secure platform is a more compelling proposition to potential users.  
    \item Token success: All else being equal, higher staking rewards increase the demand for staking. The result is an increased staking ratio and reduced liquidity in the market, which pushes token prices up. \cite{Cong} However, if staking rewards are funded by inflation and there is no additional deflationary mechanism in place, the supply of tokens may increase in the long run, putting downward pressure on the token price.
\end{enumerate}

To effectively manage the trade-offs outlined above, many blockchains allow staking reward rates to adjust dynamically. Such policies recognize that as staking rates increase, marginal gains in security decrease. In addition, the reduced market liquidity implied by higher staking ratios may hinder platform activity. A simple approach is to set an aggregate reward amount, which is then distributed to validators within a specified time frame. This allows the reward per staked token to decrease with the amount of tokens staked. While the intricacies of the policies employed vary, all of the blockchains in our sample employ staking reward policies that result in a negative correlation between the staking ratio and individual-level staking returns. 

An important additional aspect of any staking reward policy is managing the associated monetary policy. If high staking reward rates are funded through an inflationary monetary policy, token holdings are diluted and negative pressure is placed on the token price. Alternatively, high individual rewards may be financed with high transaction fees, which in turn increase entry barriers for potential users. For many Proof-of-Stake blockchains in the early stages of adoption, high inflation rates may be necessary to achieve staking rewards large enough to incentivize staking. As the platform grows and revenues from transaction fees increase, it may become possible to reduce inflation in favour of funding rewards through transaction fees. In the long run, such a transition may be necessary for token success\cite{Kose2021}.

\begin{figure}[ht]
\centering
\includegraphics[width=5in]{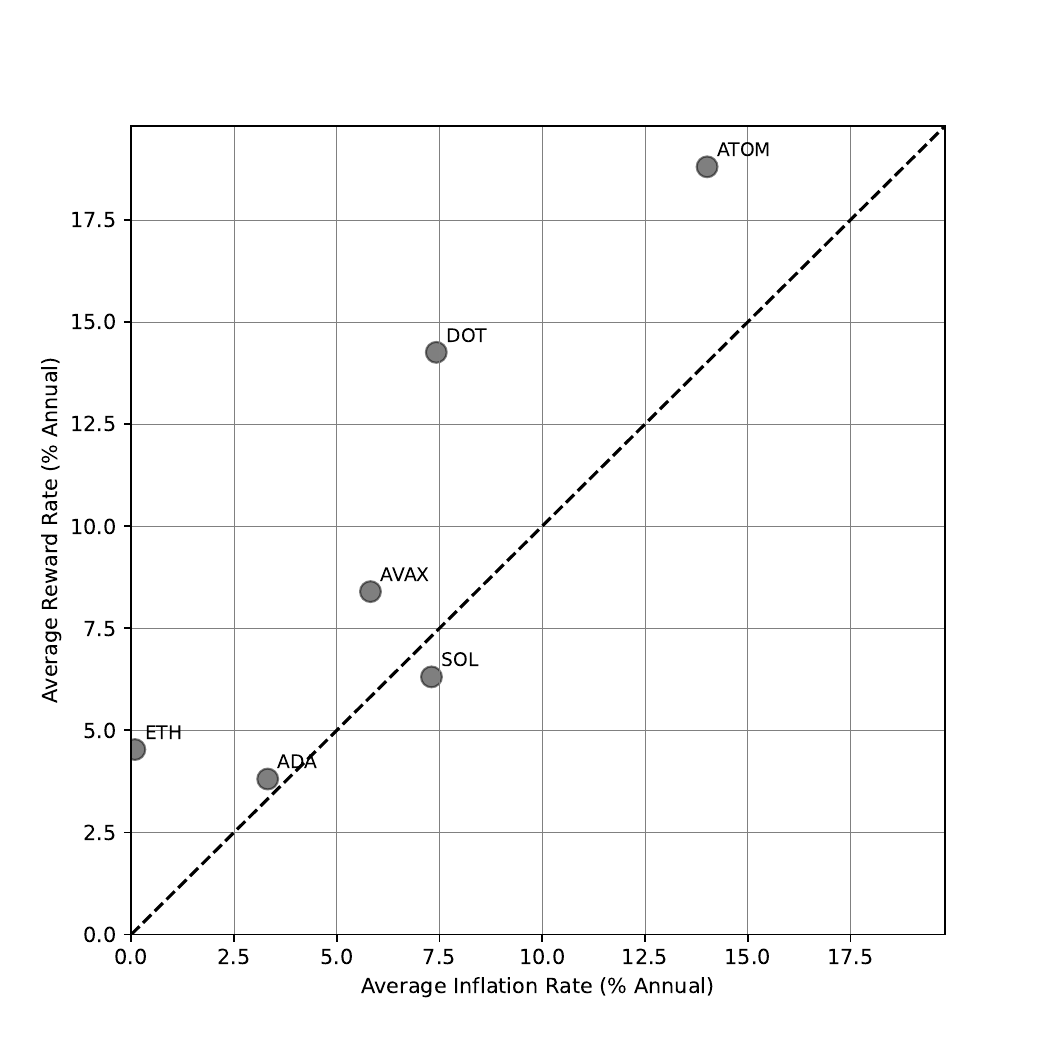}
\caption{Average Inflation Rate vs. Average Reward Rate, 2021-2023, Source: \textit{www.stakingrewards.com}}
\label{fig:2}
\end{figure}

\cref{fig:2} in \appref{appx_d} shows how average staking reward rates compare to inflation rates in the past years and is thereby illustrative of where the various blockchains stand regarding the transition to a sustainable and non-inflationary staking reward policy. First, one can observe a large variation in average nominal staking reward rates. Interestingly, protocols with longer minimum staking periods tend to also offer higher returns to staking, suggesting that stakers require a premium for the elevated price risk and losses in transaction convenience that come with longer minimum staking periods. Also, inflation and nominal staking rewards tend to be correlated, suggesting that real staking returns turn out to be much more similar and overall less attractive across blockchains. 

However, it should also be noted that many of the blockchains examined in this paper have mechanisms in place to reduce inflation as they mature. For example, Avalanche burns all transaction fees, resulting in a deflationary effect that may be able to reduce inflation as the platform grows. Although not burning transaction fee revenues, Cardano has committed to an inflation rate schedule that promises declining inflation rates over time. Solana employs a mixture of these two strategies, committing to a declining inflation rate schedule, as well as burning 50\% of the transaction fee revenues. Ethereum also follows this combined approach and already now burns enough transaction fees to allow for positive staking reward rates without inflation.

All in all, mechanisms to reduce inflationary staking reward policies are in place on most blockchains. However, whether the platforms will mature enough to allow for platform revenues to replace inflation remains to be seen. In the meantime, holders of the respective native tokens should be aware that their token holdings are at risk of substantial dilution over time.

\subsection{Slashing}
In most consensus mechanisms that we examine, various types of misbehaviour by validators can result in slashing penalties, whereby either staking rewards are reduced or the staked collateral is confiscated. Malicious behaviour can take many forms, and pinpointing individual validator actions as unambiguously malicious is not always possible. 

Certain forms of liveness attacks, for example, involve validators conspiring to cease participation in the consensus mechanism \cite{Deirmentzoglou}. Given that mere downtime does not inherently imply malicious intent, it can be difficult to identify the validators directly involved in such an attack. This can complicate the design of an effective slashing policy and present a risk for honest validators.

Slashing policies can again be evaluated in terms of the trade-off between (1) static security and (2) dynamic security: 
\begin{enumerate}
    \item Static security: Since it makes efforts to manipulate the consensus mechanism more costly, slashing can create powerful incentives for protocol-concordant behaviour. Financial losses can occur if a validator acts maliciously and is discovered. As a result, given a static set of validators, slashing increases the security of the protocol.
    \item Dynamic security: As noted above, certain validator activities, while not inherently malicious, may still be susceptible to slashing in some cases. Consequently, slashing poses a potential risk even to honest validators, and stricter slashing policies may thus act as a barrier to entry for validators wishing to join the platform. Thus, all else being equal, slashing reduces the number of validators and thus reduces security.
\end{enumerate}

As a result, slashing has a positive effect on static security and a negative effect on dynamic security. The slashing policy that maximizes security must therefore again strike a balance between the two. To find the optimal level, protocols must analyze the response of validators to changes in slashing. The lower the response of validators to slashing levels, the less important the static effect, and thus the closer the optimum will be to maximizing static security. In order to set optimal slashing parameters, protocols must therefore evaluate the slashing elasticity of staking demand.


In practice, the severity of slashing punishments and the items subjected to slashing—whether potential rewards or actual collateral—vary among blockchains. On Avalanche, rewards can be reduced if a validator’s uptime is less than 80\% of the other validators’ uptime. The protocol does not slash the collateral itself. Similarly, Cardano does not slash the collateral, but reduces staking rewards if validator pools reach a certain size threshold or if a validator does not pledge enough own stakes to their validator pool. This promotes decentralization and makes potential bribery attacks more difficult to coordinate. Finally, Algorand does not slash staked tokens but reduces rewards if token holders do not participate in the governance of the platform. 

On the remaining blockchains we evaluate, stakes tend to be slashed more severely, thus making deviations from platform-concordant behaviour more costly. On Ethereum, Solana, Polkadot, and Cosmos, the staked collateral can at least partially be slashed if the validator misbehaves. The share of tokens that are slashed can vary depending on the specific circumstances of the offense and the security risk associated with the offense.

In conclusion, we have found various trade-offs when faced with questions about security aspects. However, it remains unclear how strong these trade-offs are. Therefore, in the next section we perform a numerical analysis. In particular, we investigate whether and how factors that increase static security negatively affect dynamic security.

\section{Quantitative Assessment} \label{sec:QuantAssess}

The current section provides preliminary empirical evidence for the intuitions previously outlined. First, the data source used in the analysis is detailed, then the estimation model is described, and finally, the estimation results are presented and discussed.

\subsection{Data}

The data used in this analysis was obtained from \textit{stakingrewards.com}. The sample consists of daily observations of various staking parameters for the Ethereum, Solana, Polkadot, Cardano, Avalanche, and Cosmos blockchains over a two-year period, from 1 January 2022 to 31 December 2023. Due to concerns around data quality, we do not include Algorand in the following analysis. In addition, data on the staking parameters for each blockchain over the same two-year period was collected through desk research. The sources of these data are primarily the websites of the respective blockchains. Finally, some model specifications incorporate data on the price of Bitcoin. These data were obtained from \textit{coinmarketcap.com} and again include daily observations of the same time period. The sample of daily data is aggregated to a weekly level to reduce noise. After accounting for missing values, the sample contains 550 observations. Table \ref{sum1} in \appref{appx_d} reports summary statistics.

\subsection{Methodology}

We estimate a random-effects model with the following specification:
\begin{equation}
\Delta SR_{i,t} = \beta_{1}r_{i,t-1} + \beta_{2}\pi_{i,t-1} + \beta_{3}r_{i,t-1}\pi_{i,t-1} + \beta_{4}MA_{i,t-1} + \beta_{5} MD_{i,t-1} + \beta_{6} SD_{i,t-1} + \mathbf{\gamma} \mathbf{X} + u_{i,t}
\end{equation}

We take the first differences of the \textit{staking ratio} $SR_{i,t}$ for blockchain $i$ in week $t$ as the \textit{dependent variable}. The staking ratio $SR_{i,t}$ denotes the average proportion of the total supply of tokens staked in a given week. The first difference then represents the weekly change in the average staking ratio. The staking ratio can be seen as a proxy for the dynamic security of the protocol: as more tokens are locked into the staking protocol, validators are more strongly incentivized to contribute honestly and within the limits set by the protocol, ceteris paribus.


The first independent variables of interest in the model specification are the \textit{one-week lags of the nominal reward rate} $r_{i,t-1}$, the \textit{inflation rate} $\pi_{i,t-1}$ and their \textit{interaction}. Including an interaction term between the nominal reward rate and the inflation rate allows us to assess the interaction between rewards and inflation with regard to influencing the staking behavior. Denoted as $MA_{i,t-1}$, we also include the \textit{one-week lag of the minimum staking amount in USD} that validators must provide in order to participate in the respective consensus mechanism. In addition, we include the \textit{average minimum staking duration in days} $MD_{i,t-1}$ for blockchain $i$ in week $t-1$ and a \textit{dummy variable} $SD_{i,t-1}$ for whether validators in blockchain $i$ are exposed to having their collateral slashed or not.

Finally, several control variables are included. We include the return on the price of the native token $r_{PRICE,i,t-1}$ in the previous week $t-1$ as well as the volatility of the price of the native token $Vol_{i,t-1}$ in week $t-1$. Additionally, we include the \textit{logarithmic market capitalization} $log(Cap)_{i,t-1}$ for the native token $i$ in week $t-1$ and the \textit{logarithmic weekly trading volume} $log(Volume)_{i,t-1}$ for the blockchain $i$ in week $t-1$, as well as the return on the price of Bitcoin in week $t-1$.

\subsection{Results} 

\begin{table}
 \caption{Change in staking ratio $SR_{i,t}$ with respect to staking design parameters} \label{reg2}
 \begin{center}
{
\def\sym#1{\ifmmode^{#1}\else\(^{#1}\)\fi}
\begin{tabularx}{\textwidth}{@{\extracolsep{\fill}} l*{5}{c}}
\hline
                    &\multicolumn{5}{c}{$\Delta SR_{i,t}$}\\
                    &\multicolumn{1}{c}{(1)}&\multicolumn{1}{c}{(2)}&\multicolumn{1}{c}{(3)}&\multicolumn{1}{c}{(4)}&\multicolumn{1}{c}{(5)}\\
\hline
$r_{i,t-1}$      &       0.006         &       0.090         &       0.077         &       0.042         &       0.044         \\
                    &     (0.018)         &     (0.076)         &     (0.053)         &     (0.049)         &     (0.048)         \\
\addlinespace
$\pi_{i,t-1}$   &      -0.039         &       0.181\sym{*}  &       0.240\sym{*}  &       0.211         &       0.211         \\
                    &     (0.057)         &     (0.108)         &     (0.125)         &     (0.134)         &     (0.133)         \\
\addlinespace
$r_{i,t-1}\pi_{i,t-1}$&       0.002         &      -0.006         &      -0.006         &      -0.004         &      -0.004         \\
                    &     (0.002)         &     (0.005)         &     (0.004)         &     (0.005)         &     (0.005)         \\
\addlinespace
$MD_{i,t-1}$     &                     &      -0.080         &      -0.077\sym{*}  &      -0.073\sym{*}  &      -0.073\sym{*}  \\
                    &                     &     (0.048)         &     (0.042)         &     (0.040)         &     (0.040)         \\
\addlinespace
$SD_{i,t-1}$ &                     &      -0.166         &      -0.284\sym{**} &      -0.243\sym{*}  &      -0.245\sym{*}  \\
                    &                     &     (0.108)         &     (0.129)         &     (0.136)         &     (0.133)         \\
\addlinespace
$MA_{i,t-1}$  &                     &      -0.004         &      -0.005\sym{*}  &      -0.006\sym{**} &      -0.006\sym{**} \\
                    &                     &     (0.003)         &     (0.003)         &     (0.003)         &     (0.003)         \\
\addlinespace
$SR_{i,t-1}$    &      -0.003         &      -0.071\sym{*}  &      -0.077\sym{*}  &      -0.082\sym{**} &      -0.082\sym{**} \\
                    &     (0.004)         &     (0.040)         &     (0.041)         &     (0.039)         &     (0.038)         \\
\addlinespace
$r_{PRICE,i,t-1}$     &                     &                     &       0.006         &       0.004         &       0.002         \\
                    &                     &                     &     (0.004)         &     (0.005)         &     (0.008)         \\
\addlinespace
$Vol_{i,t-1}$       &                     &                     &      -0.001         &      -0.003         &      -0.003         \\
                    &                     &                     &     (0.002)         &     (0.003)         &     (0.003)         \\
\addlinespace
$log(Cap)_{i,t-1}$    &                     &                     &       0.187\sym{**} &       0.050         &       0.043         \\
                    &                     &                     &     (0.092)         &     (0.139)         &     (0.144)         \\
\addlinespace
$log(Volume)_{i,t-1}$&                     &                     &                     &       0.138         &       0.150         \\
                    &                     &                     &                     &     (0.090)         &     (0.098)         \\
\addlinespace
$r_{BTC,i,t-1}$     &                     &                     &                     &                     &       0.005         \\
                    &                     &                     &                     &                     &     (0.008)         \\
\addlinespace
\hline
$N$                   &     550      &     482         &     482        &     482         &     482        \\
Overall $R^{2}$              &       0.009         &       0.040         &       0.048         &       0.050         &       0.050         \\
\hline

\end{tabularx}
}
\end{center}
\vspace{1ex}

\footnotesize \textbf{Note:} Clustered standard errors in parentheses,  $*$ \(p<0.10\), $**$ \(p<0.05\), $***$ \(p<0.01\). $SR_{i,t-1}$ refers to the average staking ratio in week $t-1$, $r_{i,t-1}$ to the average staking reward rate per annum in week $t-1$, $\pi_{i,t-1}$ to the average inflation rate per annum in week $t-1$, $MD_{i,t-1}$ to the minimum staking duration in week $t-1$, $SD_{i,t-1}$ is a dummy variable for whether the protocol includes collateral slashing in week $t-1$, $MA_{i,t-1}$ is the average minimum staking amount in USD in week $t-1$, $r_{PRICE,i,t-1}$ the weekly return on the token price in week $t-1$. $Vol_{i,t-1}$ represents the volatility of the token price in week $t-1$, $log(Cap)_{i,t-1}$ the logarithm of the tokens market capitalization in week $t-1$, and $log(Volume)_{i,t-1}$ the respective weekly trading volume of the token in week $t-1$. Finally, $r_{BTC,i,t-1}$ refers to the weekly market return of Bitcoin in week $t-1$. 

\end{table}

Table \ref{reg2} reports the estimation results of different model specifications, with the preferred model reported in column (5). Although never statistically significant, all model specifications recover a positive relationship between the nominal reward rate and the change in the staking ratio in the following week. This is consistent with the basic intuition that higher reward rates increase the demand for staking. However, further analysis is needed to substantiate this intuition. Although not statistically significant, the preferred specification recovers a positive relationship between the inflation rate and the change in the staking ratio in the following week. If substantiated by further research, this suggests that higher inflation may induce some long-term token holders to stake their token holdings in an effort to compensate for the dilution of their tokens with staking rewards. However, the presence of a negative coefficient on the interaction term between the reward rate and the inflation rate may suggest that, holding the reward rate constant, increasingly high inflation rates discourage staking, potentially in favor of other staking opportunities.

Turning to the remaining staking design parameters, we find that more restrictive staking policy choices indeed tend to reduce the share of total tokens being staked. In the preferred model specification, an increase in the minimum staking amount is associated with a decline in the staking ratios, holding everything else constant. This relationship is statistically significant at the 5\% confidence level. In addition, an increase in the minimum staking duration is associated with a decline in the staking ratio of 7.3, although only at a 10\% confidence level. By comparison, the effect of the minimum staking duration is larger in magnitude. This may suggest that the increase in price risk introduced by longer lock-up periods is more prohibitive than minimum staking amounts when it comes to new stakers entering the protocol or existing stakers increasing their stake. This may be particularly important for newer protocols that aim to encourage staking to increase the economic security of their consensus mechanism, but whose native token may have higher price volatility than the tokens of more established protocols. Finally, the possibility of collateral slashing is associated with a decline in the staking ratio, however only at the 10\% confidence level. This indicates that stakers may indeed perceive severe slashing penalties as a risk, adjusting their staking behaviour accordingly. 

Overall, the results of the empirical analysis should be seen as preliminary. Future expansions of the analysis to further POS-blockchains may help to recover more robust and statistically significant estimates that can then be used as policy recommendations. In addition, the present analysis is based on aggregate measures of staking behaviour and staking rewards. Recent research suggests that blockchains differ in the degree of fairness with which they distribute staking rewards to individual validators.\cite{spychiger} Therefore, the average reward rate may, to varying degrees, misrepresent the actual staking incentives faced by individual validators, and more detailed analysis may be useful.

\section{Conclusion} \label{sec:Conclu}

This paper explores the economic and security implications of different staking designs, providing a comprehensive overview of current staking options and the economic principles that underpin them. Our analysis focuses on seven prominent POS-blockchains: Algorand, Avalanche, Cosmos, Cardano, Ethereum, Polkadot, and Solana. We identified \textit{security}, \textit{platform success}, and \textit{token success} as the primary economic goals driving growth. To achieve these goals, POS-blockchains must connect three distinct groups: validators, users, and investors. Incentivizing these groups requires significant resources, resulting in various trade-offs. For example, to fund security, blockchains may distribute new tokens, which dilutes investors, or charge users higher fees, making the platform less attractive. Blockchains address these trade-offs in different ways. In particular, their approach to addressing the security issue differs. As such, the paper highlights a \textit{strategic divergence among blockchain networks} in terms of static and dynamic security measures. Blockchains can focus either on improving the quality of the validator set for static security, or expanding the set of validators for dynamic security. Solana, Algorand, and Cardano prioritize dynamic security with low or no slashing and minimal staking requirements. In contrast, Ethereum, Polkadot, and Cosmos take a static security approach with stricter requirements to improve node quality. Avalanche is an exception to this rule. Our empiricial analysis confirms the trade-off between static and dynamic security. Our estimates show that improving the quality of the validator set, through measures such as slashing and minimum staking durations, decreases dynamic security. Overall we conclude that there is no single best staking design. The optimal design depends on specific strategies, and developers must carefully evaluate the trade-offs. To enhance understanding of the security trade-off, future research could measure validator quality (e.g. uptime) and include concentration measures in the empirical analysis.

\newpage

\ledgernotes



\begin{thebibliography}{99} \label{sec:References}

    \bibitem{Beaconcha}
    Beaconcha.in. ``Staked ETH." 
    \textit{beaconcha.in} (accessed 10 February 2024) https://beaconcha.in/charts/staked\_ether
    
    \bibitem{Saleh}
    Saleh, F. ``Blockchain without Waste: Proof-of-Stake." \textit{The Review of Financial Studies} \textbf{34(3)} 1156–1190 (2020) https://doi.org/10.1093/rfs/hhaa075.

    \bibitem{Schaaf}
    Schaaf, P., Rezabek, F., Kinkelin, H. ``Analysis of proof of stake flavors with regards to the scalability trilemma." \textit{Network} \textbf{63} (2021)

    \bibitem{Huang2021}
    Huang, H., Kong, W., Zhou, S., Zheng, Z., Guo, S. ``A survey of State-of-the-Art on Blockchains." \textit{ACM Computing Surveys} \textbf{54(2)} 1–42 (2021) https://doi.org/10.1145/3441692.

    \bibitem{Pierro&Tonelli}
    Pierro, G. A., Tonelli, R. ``Can Solana be the Solution to the Blockchain Scalability Problem?" \textit{2022 IEEE International Conference on Software Analysis, Evolution and Reengineering (SANER)} (2022) https://doi.org/10.1109/saner53432.2022.00144.

    \bibitem{Iqbal&Matulev}
    Iqbal, M., Matulevicius, R. ``Exploring Sybil and Double-Spending Risks in Blockchain Systems." \textit{IEEE Access} \textbf{9} 76153–76177 (2021) https://doi.org/10.1109/access.2021.3081998.

    \bibitem{Karpinski}
    Karpinski, M., Kovalchuk, L., Kochan, R., Oliynykov, R., Rodinko, M., Wieclaw, Ł. ``Blockchain Technologies: Probability of Double-Spend attack on a Proof-of-Stake consensus." \textit{Sensors} \textbf{ 21(19)} 6408 (2021) https://doi.org/10.3390/s21196408.

    \bibitem{Kogan2023}
    Kogan, L., Fanti, G., Viswanath, P. ``Economics of Proof-of-Stake payment Systems." \textit{Social Science Research Network} (2023) https://doi.org/10.2139/ssrn.4320274.

    \bibitem{Sai2021}
    Sai, A. R., Buckley, J., Fitzgerald, B., Gear, A. L. ``Taxonomy of centralization in public blockchain systems: A systematic literature review." \textit{Information Processing and Management} \textbf{58(4)} 102584 (2021) https://doi.org/10.1016/j.ipm.2021.102584.

    \bibitem{Irresberger}
    Irresberger, F., John, K., Saleh, F. ``The Public Blockchain Ecosystem: An Empirical analysis." \textit{Social Science Research Network} (2020) https://doi.org/10.2139/ssrn.3592849.

    \bibitem{Chitra}
    Chitra, T. ``Competitive equilibria between staking and on-chain lending." \textit{arXiv (Cornell University)} (2019) https://doi.org/10.48550/arxiv.2001.00919.

    \bibitem{Kose2021}
    John, K., Rivera, T. J., Saleh, F. ``Equilibrium staking levels in a Proof-of-Stake blockchain." \textit{Social Science Research Network} (2021) https://doi.org/10.2139/ssrn.3965599.
    
    \bibitem{Gersbach}
    Gersbach, H., Mamageishvili, A., Schneider, M. ``Staking pools on blockchains." \textit{arXiv (Cornell University)} (2022) https://doi.org/10.48550/arxiv.2203.05838.

    \bibitem{Gogol}
    Gogol, K., Kraner, B., Schlosser, M., Yan, T., Tessone, C. J., Stiller, B. ``Empirical and theoretical analysis of liquid staking protocols. " \textit{arXiv (Cornell University)} (2024) https://doi.org/10.48550/arxiv.2401.16353.

    \bibitem{Cong2022}
    Cong, L. W., He, Z., Tang, K. ``Staking, token pricing, and crypto carry." \textit{Social Science Research Network} (2022) https://doi.org/10.2139/ssrn.4059460.

    \bibitem{Noh2023}
    Noh, J., Kwon, D., Cho, S., \& Yiu, N. C. K. ``Network Participation and Accessibility of Proof-of-Stake (POS) Blockchains: A Cross-platform Comparative analysis." \textit{arXiv (Cornell University)} (2023) https://doi.org/10.48550/arxiv.2305.13259.

    \bibitem{footnote0}
    At this point, it may be important to note that several terms have evolved to describe the various possible modifications to the staking mechanism design (e.g. Delegated-Proof-of-Stake, Nominated-Proof-of-Stake, Hybrid-Proof-of-Stake, etc.). In the following, we will use Proof-of-Stake as an umbrella term to encompass these more specialized terminologies, and then introduce the modifications to the standard framework where relevant.

    \bibitem{Hafner2023}
    Hafner, M., \textit{et al.} ``DeFi Lending Platform Liquidity Risk: The Example of Folks Finance." \textit{The Journal of the British Blockchain Association} \textbf{6.1} 41-46 (2023) https://doi.org/10.31585/jbba-6-1-(5)2023.

    \bibitem{Rochet2003}
    Rochet, J.-C., Tirole, J. ``Platform Competition in Two-sided Markets." \textit{Journal of the European Economic Association} \textbf{1.4} 990–1029 (2003) http://www.jstor.org/stable/40005175.

    \bibitem{Belleflame2021}
    Belleflamme, P., Peitz, M. \textit{The Economics of Platforms: Concepts and Strategy.} Cambridge: Cambridge University Press (2021).

    \bibitem{Forbes}
    Bambisheva, N. ``Ethereum Gears Up For Next Big Upgrade; \$29 Billion Of Ether To Be Unlocked." 
    \textit{Forbes} (accessed 24 January 2024), https://www.forbes.com/sites/ninabambysheva/2023/02/09/ethereum-gears-up-for-next-big-upgrade-29-billion-of-ether-to-be-unlocked/
    
    \bibitem{Medium_StaFa}
    Staking Facilities. ``What is Nominated Proof-of-Stake?" 
    \textit{Medium} (accessed 24 January 2024) https://stakingfac.medium.com/what-is-nominated-proof-of-stake-889fc22bef8f
    
    \bibitem{Airdrops}
    No Author. ``Airdrops for ATOM holders." 
    \textit{airdrops.io} (accessed 24 January 2024) https://airdrops.io/atom-holders/
    
    \bibitem{Schmiedl}
    Schmiedl, M. ``How to stake ETH: The ultimate Ethereum 2.0 staking guide." 
    \textit{Staking Rewards} (accessed 24 January 2024) https://www.stakingrewards.com/journal/ultimate-ethereum-2-0-staking-guide/
    
    \bibitem{Cardanians}
    Cardanians.io. ``Cardano Staking: Practical Information." 
    \textit{Cardanians.io} (accessed 24 January 2024) https://cardanians-io.medium.com/cardano-staking-practical-information-3c86cbc73bd4
    
    \bibitem{Blackdaemons}
    Blockdaemon. ``Websocket Support for Ethereum." 
    \textit{Blockdaemon} (accessed 24 January 2024) https://blockdaemon.com/documentation/guides/the-ultimate-guide-to-ethereum-2-0/
    
    \bibitem{Exodus}
    Exodus. ``How do I stake ALGO with Algorand's Community Governance program?" 
    \textit{Exodus} (accessed 24 January 2024) https://support.exodus.com/article/1643-algorand-decentralized-governance-model\#
    
    \bibitem{Boshoff}
    Boshoff, K. ``How to Stake Algorand (ALGO)?" 
    \textit{Staking Rewards} (accessed 24 January 2024) https://www.stakingrewards.com/journal/how-to-stake-algorand-algo/
    
    \bibitem{AlgorandFoundation}
    Algorand Foundation. ``Algo Tokenomics." 
    \textit{Algorand Foundation} (accessed 24 January 2024) https://www.algorand.foundation/tokenomics
    
    \bibitem{AlgorandFoundation2}
    Algorand Foundation. ``FAQ - Rewards." 
    \textit{Algorand Foundation} (accessed 24 January 2024) https://www.algorand.foundation/general-faq\#04-faq-gov
    
    \bibitem{StakingRewards_AVAX}
    No Author. ``Stake AVAX Overview." 
    \textit{Staking Rewards} (accessed 24 January 2024) https://www.stakingrewards.com/asset/avalanche
    
    \bibitem{AvaLabs}
    Ava Labs. ``What is Staking?" 
    \textit{Ava Labs} (accessed 24 January 2024) https://docs.avax.network/nodes/validate/staking
    
    \bibitem{Avalanche}
    Avalanche. ``Validator FAQ." 
    \textit{Avalanche} (accessed 24 January 2024) https://support.avax.network/en/articles/6187511-validator-faq
    
    \bibitem{Avalanche2}
    Avalanche. ``What are validator/staking rewards?" 
    \textit{Avalanche} (accessed 24 January 2024) https://support.avax.network/en/articles/4587396-what-are-validator-staking-rewards
    
    \bibitem{Polkawiki}
    Polkadot. ``Introduction to staking." 
    \textit{Polkadot} (accessed 24 January 2024) https://wiki.polkadot.network/docs/learn-staking

    \bibitem{Polkadot}
    Polkadot. ``Nomination Pools are Live! Stake Natively with Just 1 DOT." 
    \textit{Polkadot} (accessed 24 January 2024) https://polkadot.network/blog/nomination-pools-are-live-stake-natively-with-just-1-dot/

    \bibitem{Polkadot_faststaking}
    Polkadot. ``Introduction to staking: Fast staking." 
    \textit{Polkadot} (accessed 24 January 2024) https://wiki.polkadot.network/docs/learn-staking\#fast-unstake
    
    \bibitem{Polkawiki_advanced}
    Polkadot. ``Advanced Staking Concepts."
    \textit{Polkadot} (accessed 24 January 2024) https://wiki.polkadot.network/docs/learn-staking-advanced\#inflation

    \bibitem{github}
    No Author. ``Validators FAQ." 
    \textit{github} (accessed 24 January 2024) https://github.com/cosmos/cosmos/blob/master/VALIDATORS\_FAQ.md
    
    \bibitem{Great_description}
    A great description of Polkadot’s staking design and reward mechanism can be found here: https://swiss-staking.medium.com/polkadot-inflation-staking-reward-4ea753380e0e
    \textit{(accessed 26 January 2024)}.
    
    \bibitem{Cosmos_Overview}
    No Author. ``Overview Cosmos Staking."
    \textit{Staking Rewards} (accessed 24 January 2024) https://www.stakingrewards.com/earn/cosmos/overview/
    
    \bibitem{Mintscan}
    No Author. ``Cosmos Overview."
    \textit{Mintscan.io} (accessed 24 January 2024) https://www.mintscan.io/cosmos
    
    \bibitem{P2P}
    P2P. ``Cosmos' (ATOM) Warm-up \& Reward Distribution Frequency."
    \textit{P2P} (accessed 24 January 2024) https://p2p.org/faq/en/articles/5286271-cosmos-atom-warm-up-reward-distribution-frequency

    \bibitem{cosmos_FAQ}
    Cosmos Hub. ``Validator FAQ."
    \textit{Cosmos Hub} (accessed 29 January 2024) https://hub.cosmos.network/main/validators/validator-faq.html

    \bibitem{figment}
    Figment. ``Cosmos Staking Guide."
    \textit{Figment} (accessed 29 January 2024) https://figment.io/insights/cosmos-hub-staking-guide/
    
    \bibitem{Cardano}
    Cardano. ``Pledging and Rewards."
    \textit{Cardano} (accessed 29 January 2024) https://docs.cardano.org/learn/pledging-rewards
    
    \bibitem{coinbase}
    Coinbase. ``Guide to Cardano."
    \textit{Coinbase} (accessed 29 January 2024) https://www.coinbase.com/de/cloud/discover/protocol-guides/guide-to-cardano

    \bibitem{coinmarketcap}
    No Author. ``Overview Cardano."
    \textit{Coinmarketcap} (accessed 29 January 2024) https://coinmarketcap.com/currencies/cardano/

    \bibitem{Solana}
    Solana. ``Staking Overview."
    \textit{Solana} (accessed 29 January 2024) https://solana.com/staking\#overview/what-is-proof-of-stake 
    
    \bibitem{Solana_docs}
    Solana. ``Inflation Related Terminology."
    \textit{Solana} (accessed 29 January 2024) https://docs.solana.com/inflation/terminology
    
    \bibitem{Solana_docs2}
    Solana. ``Stake Accounts"
    \textit{Solana} (accessed 29 January 2024) https://docs.solana.com/staking/stake-accounts
    
    \bibitem{Solana_docs3}
    Solana. ``Delegation Timing Considerations."
    \textit{Solana} (accessed 29 January 2024) https://solana.com/de/staking\#overview/delegation-timing-considerations

    \bibitem{Solana_docs4}
    Solana. ``Proposal Optimistic Confirmation and Slashing."
    \textit{Solana} (accessed 20 April 2023) https://docs.solana.com/proposals/optimistic-confirmation-and-slashing

    \bibitem{Solana}
    Solana. ``Staking Overview."
    \textit{Solana} (accessed 29 January 2024) https://solana.com/staking\#overview/what-is-proof-of-stake
    
    \bibitem{Ethereum_book}
    Edgington, B. 
    \textit{Upgrading Ethereum: A technical handbook on Ethereum's move to proof of stake and beyond.} No city: No Publisher 100-108 (2023).

    \bibitem{Ethereum_book2}
    Edgington, B. 
    \textit{Upgrading Ethereum: A technical handbook on Ethereum's move to proof of stake and beyond.} No city: No Press 92-96 (2023).

    \bibitem{Ethereum_book3}
    Edgington, B. 
    \textit{Upgrading Ethereum: A technical handbook on Ethereum's move to proof of stake and beyond.} No city: No Press 115-119 (2023).

    \bibitem{Coindesk2}
    Nijkerk, M. ``Ethereum’s Shanghai Hard Fork Now Has Official Target Date."
    \textit{Coindesk} (accessed 29 January 2024) https://www.coindesk.com/tech/2023/03/16/ethereums-shanghai-hard-fork-now-has-an-official-target-date/

    \bibitem{Lutz}
    Lutz, S. ``Proposal Optimistic Confirmation and Slashing."
    \textit{Decrypt} (accessed 29 January 2024) https://decrypt.co/125186/ethereum-shanghai-upgrade-means-you-eth-sec

    \bibitem{footnote}
    Solana allows no more than 25\% of the total active stake on the network to be deactivated in a single epoch. Ethereum implements a similar limit, which can be adjusted depending on the size of the active validator set.
    
    \bibitem{Cong}
    Cong, L. W., He, Z., Tang, K. ``The Tokenomics of Staking." 
    \textit{FTG Working Paper Series} \textbf{00134-00} (2023) http://dx.doi.org/10.2139/ssrn.4320274.
    
    \bibitem{Deirmentzoglou}
    Deirmentzoglou, E. \textit{et al.} ``A Survey on Long-Range Attacks for Proof of Stake Protocols." 
    \textit{IEEE access} \textbf{7} 28712-28725 (2019) 0.1109/ACCESS.2019.2901858.
    
    \bibitem{footnote_appendix}
    All prices in USD are according to the prices as of the 4th of March, 2024.

    \bibitem{spychiger}
    Li, S. N., Spychiger, F., \& Tessone, C. J. (2023). Reward Distribution in Proof-of-Stake Protocols: A Trade-off Between Inclusion and Fairness. IEEE Access, 11, 134136-134145.
    
\end{thebibliography}

\newpage

\appendix

\setcounter{section}{0} 
\renewcommand{\thesection}{\Alph{section}}
\section{Different Ways of Staking} \label{appx_a}

\subsection{Variations of POS} 
In a classical POS system, such as Ethereum, stakers provide the stake and the computational power necessary to verify and approve the next block of transactions. In these systems, only the so-called validators can stake. In other words, the ownership of the invested capital and the executive function it grants are both retained by the same agent. Over time, permutations of the POS mechanism have been developed that allow the separation of validation and staking, attributing separate roles to validators and stakers.

The simplest advancement of POS is the Delegated-Proof-of-Stake (DPOS). In DPOS-blockchains, like the Avalanche Network, a delegator can commit his stake to the validator. The validator, in turn, operates a staking pool that collects the assets from the various delegators. As in the classical POS mechanism, validators also have the task of appending new blocks to the blockchain. Staking rewards are distributed among the validators and the delegators as well.

Another extension of POS is the Nominated-Proof-of-Stake (NPOS) mechanism that was first implemented by Polkadot. Similar to DPOS the roles of staking and validation are separated. Validators are elected by the so-called nominators to participate in the block-update process. Compared to DPOS, however, a nominator's stake is allocated by a predefined algorithm among those validators who had received the most votes. Once an active set of validators is established by the voting result, the staked assets are distributed among the validators to ensure sufficient decentralisation.\cite{Medium_StaFa} Importantly, this consensus mechanism introduces reputation as a defining attribute of validator success. Nominators will only commit their assets to validators whom they trust to act truthfully and in a manner that will maximise their staking returns. 
Compared to DPOS, NPOS has introduced a new role for the delegators/nominators. Nominators get the additional role of nominating validators in addition to providing financial backing. The nominator stakes are then distributed among a set of active validators, unlike in DPOS, where stakes are delegated to a single validator of choice. 
\subsection{Different Ways of Staking}
Participating in POS as a delegator or validator usually requires technical knowledge to mitigate the risk of slashing. Since most individuals do not have sufficient knowledge, third parties (such as Lido or Coinbase) started to offer staking as a service. These service providers allow anyone to participate in the staking process without any prerequisites. We broadly delineate the staking possibilities and describe their potential risks and benefits among two dimensions:
\begin{enumerate}
    \item whether individual stakers stake directly on the blockchain itself or through a third-party platform (indirectly) and
    \item if the third-party platform provides a custodial or a noncustodial service.
\end{enumerate}
\subsection{Direct vs. indirect staking:}
Various definitions of direct staking exist. We define it as the act of individuals directly staking their assets on a POS-blockchain, which consequently positions them as active participants within the POS ecosystem. In comparison, we define indirect staking as when users stake through a third-party and are itself not a participant of the POS mechanism. Note, we consider delegators and nominators as direct stakers since they lock their assets on the POS-blockchain by themselves and not through a third-party and are a primary agent of the POS ecosystem.

Direct staking can be conducted using an own validator or by delegating/nominating stakes. On most blockchains, running a validator node requires having access to hardware infrastructure, knowledge of a specific software environment, and active participation in the consensus mechanism of the blockchain. While computing power requirements are much less prohibitive than on POW protocols, successful validator nodes still need to have access to a stable internet connection and be able to run their node perpetually and without interruptions. A less resource-intensive form of direct staking is when individuals stake their own assets directly as part of the POS system on the blockchain as a delegator/nominator. 

While being the most involved form of staking, one of the major benefits of direct staking is that one directly accrues potential staking rewards. Above that, on some blockchains, like Cosmos, direct stakers regularly receive airdrops.\cite{Airdrops} Fees for direct staking are lower in the case of delegating/nominating or non-existent in the case of running an own validator in comparison to staking through a third-party. Direct staking requires individuals to have some knowledge of blockchain technology whilst indirect does not. Overall, the trade-off between direct and indirect staking tends to be determined by the risk associated with the transfer of ownership and the search and transaction costs associated with staking through a third-party.

\subsection{Custodial vs. non-custodial staking:}
Indirect staking can be custodial or non-custodial, depending on whether the staker remains the owner of their assets (non-custodial) or transfers the ownership to a third-party platform (custodial). In custodial staking, the fact that ownership of the staked assets shifts from the staker to the custodian represents a risk that is not present when a staker stakes through a non-custodial platform or directly on the blockchain. On the other hand, custodial staking is the easiest way to start staking. Custodial staking service providers may, for example, offer staking with no lock-up period or no minimum required amount, which are often unavoidable in other forms of staking. 
On non-custodial staking platforms, delegators retain possession of their private key and hence their staked coins.\cite{Schmiedl} This form of staking mitigates the intermediary in comparison to the custodian form. In addition, staking rewards tend to be higher on non-custodial platforms.

\section{Staking Characteristics / Dimensions}\label{appx_b}
Beyond the general concept of staking and its different forms, it is important to understand some of the common dimensions along which blockchains define their Proof-of-Stake mechanisms. This is the purpose of this section. Furthermore, based on the characteristics described here, a detailed and comparative overview of the most prominent blockchains are to follow in the next Appendix.

\begin{enumerate}
\item Epoch: The epoch is usually defined as a period of time or a specific number of blocks. During an epoch, blocks are submitted, validated and at the end staking rewards are distributed. The length of an epoch varies considerably by blockchain. For instance, epochs on Cardano extend over a period of 5 days \cite{Cardanians}, whereas Ethereum has 6.4-minute-long epochs.\cite{Blackdaemons}
\item Aggregated rewards: Aggregated rewards are the staking rewards on a blockchain that are to be distributed among all stake holders. Aggregated rewards are calculated differently on each blockchain and can depend on different variables, like inflation or the share of total staked coins in the network.
\item Staking pool: Direct stakers may operate staking pools through which indirect stakers can provide their coins. The pool represents the combined stakes. Indirect stakers receive a share of the rewards earned by the pool.
\item Slashing: Validators should be online all the time and avoid any misbehaviour to ensure network security and stability. However, if they have problems with their infrastructure and go offline or double-sign a block, the stakes in their staking pool can get slashed. Alternatively, aggregated rewards are reduced.
\item Individual rewards: Individual rewards are distributed among individual stakers. They usually depend on the individual stake made by the staker and can be reduced by any validator and/or third-party fees and slashing penalties.
\item Unbonding period: Some blockchains impose an unbonding period on stakers. The unbonding period starts when a staker initializes the withdrawal of their stakes. During the unbonding period, the coins are locked on the blockchain, and their owner cannot use these in any transaction or earn further staking rewards. Unbonding periods can be a few hours or even many weeks. 
\item Activation period: On some blockchains, staked coins do not start earning rewards as soon as they are staked. Often, users may have to wait until the beginning of a new epoch for their stake to start earning staking rewards. The activation period then refers to the period between the staking commitment of the user and the integration of the stake into the active Proof-of-Stake consensus mechanism.
\item Compounding rewards: Some blockchains automatically distribute the earned rewards into the stock of staked coins, thereby increasing the number of staked tokens and allowing the staker to benefit from the compound interest effect. Reward distribution and crediting are not always automatic though. In some cases, they must be initialized by the validator.
\end{enumerate}

\section{Overview of the POS Implementations} \label{appx_c}
In this section of the appendix, we are going to describe the staking mechanism of seven prominent blockchains. The goal of this part is to give a summary of our findings and of the mechanisms on which our analysis is based. Even though, we updated these information during our whole research process several times, the descriptions of some staking designs are scarce and could have changed since our last update.

\textit{Algorand} --- The first Blockchain we discuss is Algorand which implemented a major change in its initial staking design. In the beginning, every ALGO holder automatically earned staking rewards. In the new governance program, ALGO holders receive their staking rewards only if they become governors and meet the following two conditions. First, they must provide their stakes for at least three months. Second, they have to vote on initiatives about the growth and development of the Algorand ecosystem.\cite{Exodus} The staked coins in the wallet of a governor are not locked, which means that they can make transactions with those coins. However, whenever the number of coins stored in the wallet falls below the committed number of coins, the governor becomes ineligible for rewards in that governance period. There is no minimum amount of ALGO that has to be staked, but it should be larger than the transaction fees.\cite{Boshoff} Furthermore, governors are required to vote on every proposal, if they want to earn the maximum number of coins, otherwise, their rewards will be slashed. Uniquely, if a governor misses the sign-up period lasting for two weeks at the beginning of a governance period, they will only be able to participate in the next governance period, in three months.

From the 10 bn initially minted ALGOs, 6.9 bn are already in the circulating supply and the remaining 3.1 bn is held by the Algorand Foundation.\cite{AlgorandFoundation} The aggregated reward pool is funded with the 3.1 bn stock of ALGO held by the Algorand Foundation. The individual rewards of the governors depend on the ratio of the total stakes committed and the total rewards for the given period.\cite{AlgorandFoundation2} Since the distribution of individual rewards for each governor depends on their participation, rewards are not compounding and are not continuously accounted to a wallet. Governors receive their rewards at the end of the 3-months governance period.

Compared to the previous staking mechanism, in which every ALGO holder could earn staking rewards, this mechanism is more restrictive. If a potential governor misses the sign-up period, he is completely excluded from the governance period. It also requires an active willingness to stake, but this results in a trade-off of less decentralization. 

The new mechanism in which Algorand eliminates the rewards as a penalty of not participating in governance can be seen as a slashing mechanism. In its initial, passive staking mechanism, there was no similar slashing mechanism in place. The lack of such led to criticism and was seen as a potential source of risk because misbehaving nodes had little to lose.

\textit{Avalanche} --- The next blockchain in our analysis, Avalanche, has a completely different staking mechanism compared to Algorand. Stakers are not required to be governors and to vote on proposals. Coin holders must delegate their stake to a validator of their choice or become a validator themselves. The total supply of AVAX is capped at 720 mio. Initially 360 mio. AVAX were minted. Staking rewards have been continuously minted since then and as the circulating supply gets closer to the cap, the reward rates are being reduced.

Individual rewards depend on the total amount staked on the network and the duration of the stake.\cite{StakingRewards_AVAX} The minimum staking period should last at least two weeks, and the maximum can last up to one year. Depending on the staking period, the current annual reward rates are 7.9\% and 9.4\%, respectively. The minimum number of AVAX that must be delegated for staking is 25 AVAX. The minimum number of AVAX a validator who sets up the staking pool must stake is 2000. Unlike on Algorand and other chains, the maximum number of AVAX a validator is allowed to collect into a single staking pool can only be 5 times his own stake. This means that if a validator stakes 1000 AVAX, he can only take a total of 4000 AVAX from delegators and has a total of 5000 AVAX in his pool.\cite{AvaLabs} The maximum amount of AVAX one can stake is 3 millions.\cite{Avalanche}

A noteworthy similarity to Algorand is that in their documentation they do not mention slashing as such. However, if a validator is online less than 80\% of the time most validators are online, she does not receive any rewards.\cite{Avalanche2} Rewards are paid only after the committed staking period is over, therefore stakes and rewards do not compound.\cite{AvaLabs} On the Avalanche network stakes are locked up for the committed staking period. This carries an additional risk to stakeholders because, in case of an AVAX devaluation, the staked coins could devaluate as well and cannot be sold on the market.

\textit{Polkadot} --- This blockchain took up the mission of solving the interoperability problem. To tackle the problem, it connects several chains together through so-called shards. This allows parallel transaction processing, data exchange, and the optimization of each chain on the network for a unique specification. Compared to the already discussed blockchains, Polkadot implements a novel kind of POS consensus mechanism, called Nominated Proof-of-Stake (NPOS). In this mechanism there are two essential roles, which should be elaborated on: \textit{validators} and \textit{nominators}. Validators provide the infrastructure of the network, they validate the blocks, whereas nominators contribute with their stakes and nominations to the security of the network. Nominators can suggest validators into the active set of validators and stake their tokens with the chosen validators.\cite{Medium_StaFa} As of March 2023, at least 313 DOTs are required from nominators. Through the introduction of nominator pools, a new way of staking was created, which allows for more scalability. Nominators can join a nominator pool with a stake of just 1 DOT.\cite{Polkawiki} The lower minimum stake required is not the only advantage of nomination pools. They enlarge the pool of potential stakers by lowering the barrier of entry into staking, which is created by the due diligence that should be done before staking. The nominator pool creator does all the research on behalf of other pool members.\cite{Polkadot} Figure 1 below visualizes the two different ways of staking with Polkadot. A further major difference to the other blockchains considered is, that in the case of Polkadot, nominators are completely free to choose up to 16 validators whom they are backing with an arbitrary number of coins. Polkadot then automatically chooses the validator to which the coins are allocated. This reduces the risk for the nominators of not getting rewards due to validator inactivity. 

Unlike on other networks, rewards do not depend on the share a validator has in the aggregated stakes. Rewards are only influenced by their activity on the chain, with which they can earn “era points”. Any misbehaviour, or inactivity of a validator will lead to slashing of stakes for both the validator and its nominators. A noteworthy difference in the slashing mechanism compared to others is that popular validators will be slashed proportionately more, to incentivize nominators staking with less popular validators to decrease their individual risks and increase decentralization. Any slashed stakes are transferred to the Treasury and can be reverted only by the Council. 

On Polkadot, rewards are accumulated each era, which is approx. a 24-hours-period. Rewards pay-out must be initialized by either the validator or one of the nominators and rewards can be directed to an unrelated account or to the account containing the stake. Polkadot offers a lot of flexibility when it comes to withdrawing or increasing the stakes. It is possible to increase or reduce the stake without the need to unstake all coins. On our list of blockchains, the longest unbonding period has been implemented by Polkadot. Stakes remain locked for 28 days if someone wishes to get them back.\cite{Polkawiki}

However, a new feature is coming soon, and fast-unstake is going to be added. This will allow nominators to perform unstaking, if their balance did not back any validators in the last 28 days.\cite{Polkadot_faststaking}

An important similarity to Cosmos is the dependence of aggregated rewards on the share of staked coins. On Polkadot the ideal staking rate on the network was defined to be 50\%, but this can change depending on the number of parachains. The Rewards pool is determined by the inflation rate, whereas the inflation rate depends on the system staking rate. If the system staking rate is lower than the ideal staking rate of 50\%, inflation (newly minted coins) increases linearly up to 10\%. The closer the system staking rate gets to the ideal staking rate, the higher the inflation rate and thus the higher aggregate rewards for all stakers become. If the system staking rate is larger than the ideal, the inflation rate and reward rate start to decrease in an exponentially decaying manner.\cite{Polkawiki_advanced} Therefore, the aggregated rewards are highest when the system staking rate equals the ideal staking rate. 

Individual staking rewards decrease with increasing staking rate (despite increasing total staking rewards up to the ideal staking rate). This is by the simple fact, that with more stakers, the rewards have to be distributed to more stakers. Since this effect is stronger than the increase in total rewards, individual staking rewards decrease.

\textit{Cosmos} --- The internet of blockchains, employs a POS consensus mechanism. The probability of a validator being chosen for block proposal depends on the relative size of their validator pool, to which they and delegators can contribute. Rewards are composed of so-called block provisions and transaction fees, which are distributed among validators proportionally to the size of their pool. Within the staking pool, rewards are again distributed among delegators in proportion to their individual contribution to the pool.\cite{github}
Similar to Polkadot’s reward design\cite{Great_description}, block provisions on Cosmos are inflationary and their total amount is determined by the inflation rate, which in turn depends on the share of the circulating ATOM supply being staked. The desired staking participation on the network is determined by Cosmos to be 66\%. Deviations from the target staking rate are corrected through the inflation rate. If the staking rate is less than 66\%, inflation gradually increases up to at most 20\% per annum. A higher inflation rate leads to increased staking rewards as well as to a stronger dilution of non-staked ATOM. Combined, these effects incentivize additional staking. If more than 66\% of all ATOM are staked, the policy becomes to reduce the incentives for staking by decreasing inflation gradually down to a floor of 7\% per annum.\cite{Cosmos_Overview} As of the time of writing, inflation is 16\% and the staking ratio is around 67\%.\cite{mintscan}

Cosmos implements an unbonding period of 21 days, during which assets cannot be used in transactions or earn any rewards. This unbonding period leads to a similar problem as the commitment period in the case of the Avalanche network. Locked coins are at risk of experiencing a complete value deterioration during the 21-day unbonding. An additional restraint is that unbonding stakes through the same validator is allowed only 7 times in a 21-day period, which in the short run could result in stakes being stuck with a validator. However, staking rewards are accruing every 7 seconds on a separate account and can be used in transactions or staked separately.\cite{P2P}

There is no minimum staking requirement for the delegators except for the 0.05 ATOMs (USD 0.6), which must be staked to be able to unstake the coins in the future. The minimum stake required from a validator is in theory 1 ATOM. However, if a validator’s total stake falls below the top 175, then that validator loses their right to validate.\cite{cosmos_FAQ} Therefore, the effective minimum staking requirement for a validator is much larger than the theoretically required 1 ATOM. No maximum staking amount is defined by the staking protocol. A potential downside from the perspective of an individual delegator, but a benefit to the functioning of the consensus mechanism as a whole is that stakes can get slashed. In a staking pool, 0.01\% of the stakes are slashed for the downtime of a validator, whereas 5\% are slashed for double signing activity.\cite{figment} Even stakes subject to the unbonding period can get penalized for downtime or double signing.

\textit{Cardano} --- One of the differences between Cardano and the blockchains discussed above is its mechanism to limit the amount of stakes a staking pool aggregates. If a staking pool on Cardano attracts more stakes than the threshold of 1/k, the individual rewards for the specific staking pool will no longer increase. The parameter k represents the so-called stake cap, which limits the number of ADA that can earn rewards in a stake pool. It is possible to stake ADA above the value of parameter k into the stake pool, but that would be irrational, since the additional stakes would not earn any rewards. The goal is to set k such that 1000 stake pools are created and running on the network.\cite{Cardano} The rewards of the individual staking pools also depend on whether the validator pledged own ADA to the pool. Without any pledged ADA, the specific staking pool can only earn 77\% of the rewards a similar stake pool with pledged ADA could earn. This mechanism incentivizes validators to stake their own ADA alongside delegators, which further reduces their incentives for misbehaviour.\cite{coinbase}
Another type of penalty or slashing that is applied to individual staking rewards, is the reduction of rewards when block validation is missed. However, only the rewards are slashed, not the actual stake. There is no minimum staking amount required, but at least 10 ADAs (USD 8) should be staked, to be able to finance the fee of unstaking later. Unlike Cosmos, there is no unbonding period, delegators are free to unstake whenever they wish to.

The aggregated reward pool is calculated in each epoch as follows:

First, the Reserve is calculated. The Reserve is the difference between the maximum possible supply of ADA, which was set to be 45 bn and the ADA in circulation and in the Treasury.\cite{coinmarketcap} Then, the Total Possible Rewards are calculated with the help of a parameterizable rate which was determined to be 0.3\%. A predefined rate of 20\% of the Total Possible Rewards is directed as a Treasury fee into the Treasury, and the remainder will fill the Aggregated Rewards Pool. However, the rewards distributed from this remainder are proportionate to the staking rate of the system. If only 50\% of all coins are staked, then 50\% of the Aggregated Rewards will be distributed and the other 50\% is sent back to the Treasury.\cite{Cardano}

The distribution of the remaining 50\% of the Total Possible Rewards is as follows:

The share of Aggregated Rewards a pool can get depends on numerous variables such as the Total Possible Rewards, the relative pool saturation size, the amount pledged by the pool operator, the performance of the validator, and the amount staked by delegators. The more ADA is staked in the pool, the higher the share of rewards, of course subject to the threshold mentioned above. Also, if the validator had not pledged any own coins, the share of Aggregated Rewards the pool can receive, is reduced. Before distributing the rewards to delegators, proportionally to their individual stakes, validator fees are subtracted. 

\textit{Solana} --- This blockchain combines Delegated-Proof-of-Stake and Proof-of-History consensus mechanisms to provide a highly scalable blockchain with high transaction speeds and low transaction fees. Like many other Proof-of-Stake blockchains, Solana plans to finance staking returns through inflation (issuance of their native coin SOL). The platform foresees 100\% of inflation to be distributed to validators and staking accounts. The inflation rate is set to begin at 8\% per annum and is scheduled to decrease by 15\% each year until it ultimately reaches 1.5\%.\cite{Solana}

Being a Delegated-Proof-of-Stake platform, individuals can access staking rewards by delegating SOL into validator pools. Staking rewards are distributed among validators according to their share of the total amount of SOL being staked. From the perspective of an individual delegating their SOL, the annualized staking yield can be characterized as follows.\cite{Solana_docs}

\begin{equation}
\small
{\textit{Staking Yield}}=\textit{Inflation Rate}\cdot \textit{Validator Uptime}\cdot (1-\textit{Validator Fee})\cdot\frac{\textit{Total Current Supply}}{\textit{Total SOL Staked}}   
\end{equation}

Validator uptime and validator fees vary across validators. Validator uptime is a measure of the number of votes a validator participates in during a given epoch. Thus, the more actively a validator participates in the consensus mechanism, the higher their staking rewards. In the case of Solana, one epoch has a duration of around 2 days. Additionally, delegators compensate validators in the form of a validator fee. Staking returns are paid out after every epoch and are automatically restaked. Also, stakers do not have to comply with any minimum or maximum staking requirements and can withdraw their stake on relatively short notice, at the end of an epoch.\cite{Solana_docs2} Given the duration of an epoch on the Solana blockchain, this means that the unbonding period is at most 2.5 days. There is also a similar activation period. New stakes are activated only at the beginning of a new epoch.\cite{Solana_docs3} Currently, slashing on the Solana blockchain is administered on a case-by-case basis. However, validators who sign illegal transactions or vote for invalid forks may see 100\% of their stakes slashed.\cite{Solana_docs4}

\textit{Ethereum} --- The introduction of Proof-of-Stake to the Ethereum Mainnet represented a major step in the roll-out of the new consensus mechanism on Ethereum. Of course, staking has been possible for a while on the Beacon Chain. However, we focus on the staking policies that have been developed since the successful merge of the Beacon Chain and the Ethereum Mainnet when rewards ceased to be paid out to POW miners and the full transition to POS was completed. We further discuss the changes to the Ethereum staking policy that stand to be implemented with the Shanghai/Capella-upgrade scheduled for the 12$^{th}$ of April.

Activation of a validator node in the Ethereum ecosystem requires an initial stake of at least 32 ETH. Individuals can either set up a validator node on their own, employ the services of a third-party node operator, or contribute their stake to an existing third-party staking pool. Validators running their own node can primarily earn rewards through attestation and block proposal. In each epoch, validators may attest their perspective of the chain within the context of the consensus protocol, thereby regularly earning attestation rewards. Attestation rewards increase with the validator’s so-called effective balance. Additionally, a validator may be randomly chosen by the protocol to provide a block proposal in a given epoch. Since this is a random process, a given validator will not be asked to contribute a block proposal in every epoch, making the rewards paid out for this activity more variable. The likelihood of being selected for a block proposal increases with a validator’s effective balance.\cite{Ethereum_book}

While actual staking balances fluctuate when rewards are paid out or stakes are slashed, the effective balance adjusts more slowly and is capped at 32 ETH.\cite{Ethereum_book2} This means that compounding rewards are not possible in the strict sense, because rewards accruing to the actual balance do not always affect a change in the effective balance.
Validator node downtime does not incur slashing of the underlying stake but may reduce validator rewards. On the other hand, actions such as double signing or malicious voting behaviour may result in slashing. The amount of ETH slashed depends on the number of other validators slashed at the time. Since a successful attack on the network requires a majority of ETH validators to cooperate, widespread misbehaviour is severely punished while sporadic accidental violations are treated much more leniently.\cite{Ethereum_book3}
\newpage
\section{Empirical Calculations} \label{appx_d}

\begin{table}[!htbp]
 \caption{Summary statistics, by token} \label{sum1}
 \begin{center}{\resizebox{\textwidth}{!}
 {
     \begin{tabular}{l*{7}{c}}
    \hline
    &ADA & ATOM &AVAX & DOT &ETH &SOL& Overall\\
    \hline
    $SR_{i,t-1}$& 68.12 (3.83)& 65.09 (3.09)& 60.66 (3.71)& 49.29 (3.60)& 14.29 (4.59)& 72.83 (2.61)& 54.84 (20.12)\\
    $r_{i,t-1}$& 3.83 (0.87)& 18.76 (3.10)& 8.40 (0.64)& 14.25 (0.31)& 4.54 (0.82)& 6.30 (0.73)& 9.33 (5.59)\\
    $\pi_{i,t-1}$& 3.34 (0.54)& 13.99 (3.25)& 5.82 (0.38)& 7.43 (0.36)& 0.10 (0.38)& 7.32 (0.27)& 6.17 (4.67)\\
    $MD_{i,t-1}$& 0.00 (0.00)& 21.00 (0.00)& 14.00 (0.00)& 28.00 (0.00)& 0.00 (0.00)& 5.00 (0.00)& 12.70 (10.44)\\
    $SD_{i,t-1}$& 0.00 (0.00)& 1.00 (0.00)& 0.00 (0.00)& 1.00 (0.00)& 1.00 (0.00)& 1.00 (0.00)& 0.67 (0.47)\\
    $MA_{i,t-1}$& 0.00 (0.00)& 0.00 (0.00)& 7.44 (6.80)& 0.09 (0.06)& 615.46 (202.31)& 0.00 (0.00)& 103.83 (243.35)\\
    $r_{PRICE,i,t-1}$& -0.29 (9.11)& -0.21 (9.44)& -0.01 (12.52)& -0.83 (7.98)& -0.20 (8.06)& 0.31 (13.47)& -0.20 (10.28)\\
    $Vol_{i,t-1}$& 0.02 (0.03)& 0.70 (0.77)& 1.68 (2.05)& 0.42 (0.53)& 67.83 (56.37)& 2.81 (3.43)& 12.25 (33.88)\\
    $log(Cap)_{i,t-1}$& 23.46 (0.44)& 22.05 (0.42)& 22.64 (0.62)& 22.90 (0.49)& 26.12 (0.29)& 23.27 (0.64)& 23.41 (1.39)\\
    $log(Volume)_{i,t-1}$& 21.85 (0.77)& 21.20 (0.81)& 21.55 (0.83)& 21.37 (0.77)& 25.07 (0.46)& 22.49 (0.76)& 22.26 (1.52)\\
    $r_{BTC,i,t-1}$& 0.15 (6.75)& 0.15 (6.75)& 0.15 (6.75)& 0.15 (6.75)& 0.15 (6.75)& 0.15 (6.75)& 0.15 (6.72)\\
    \hline
    \multicolumn{8}{l}{Notes: Table reports Mean (SD)}\\
    \end{tabular}
}
}\end{center}
\end{table}
\newpage


\newpage




\thispagestyle{pagelast}

\end{document}